\documentclass[article]{elsarticle}

\usepackage{amsmath}
\usepackage{amsfonts}
\usepackage{amssymb}
\usepackage{graphics}
\usepackage{graphicx}

\usepackage{multirow}

\usepackage{color} 
\definecolor{mygreen}{RGB}{28,172,0} 
\definecolor{mylilas}{RGB}{170,55,241}
\usepackage{subcaption} 
\journal{...}

\begin{document}

\begin{frontmatter}

\title{Assessment of a porous viscoelastic model for wave attenuation in ice-covered seas}

\author[a1]{Boyang Xu}
\ead{boyangxu@udel.edu}

\author[a1]{Philippe Guyenne\corref{cor1}}
\ead{guyenne@udel.edu}

\address[a1]
{Department of Mathematical Sciences, University of Delaware, DE 19716, USA}

\cortext[cor1]{Corresponding author}


\begin{abstract}
Chen et al. \cite{cgg19} recently proposed a two-dimensional continuum model for linear gravity waves propagating in ice-covered seas.
It is based on a two-layer formulation where the ice cover is viewed as a porous viscoelastic medium.
In the present paper, extensive tests against both laboratory experiments and field observations are performed
to assess this model's ability at describing wave attenuation in various types of sea ice.
The theoretical predictions are fitted to data on attenuation rate via error minimization and numerical solution
of the corresponding dispersion relation.
Detailed comparison with other existing viscoelastic theories is also presented.
Estimates for effective rheological parameters such as shear modulus and kinematic viscosity
are obtained from the fits and are found to vary significantly among the models.
For this poroelastic system, the range of estimated values turns out to be relatively narrow 
in orders of magnitude over all the cases considered.
Against field measurements from the Arctic Ocean, this model is able to reasonably reproduce 
the roll-over of attenuation rate as a function of frequency.
Given the rather large number of physical parameters in such a formulation,
a sensitivity analysis is also conducted to gauge the relevance of a representative set of them to the attenuation process.
\end{abstract}


\end{frontmatter}


\section{Introduction}

In recent decades, the polar regions have experienced major transformations due to global warming.
For example, the rapid decline of summer ice extent in the Arctic Ocean has caught a lot of attention.
While there is no doubt that warmer temperatures have been a major factor in transforming the polar seascape,
evidence has also shown that ocean waves and their increased activity play an aggravating role, 
and in turn the presence of sea ice affects the wave dynamics.
By breaking up the sea ice, waves cause it to become more fragmented, which in turn increases
their capacity to further penetrate and damage the ice cover.
A typical setting in the ocean where wave-ice interactions prevail is the marginal ice zone (MIZ)
which is the fragmented part of the ice cover closest to the open ocean.
It is a highly heterogeneous region comprising various types of sea ice that result from the incessant assault of incoming waves.

Of particular interest to oceanographers is the modeling of wave attenuation in sea ice,
a process that has been poorly represented in large-scale wave forecasting models for the polar regions.
There are two principal mechanisms for the attenuation of wave energy propagating into an ice field:
(i) scattering by ice floes or other inhomogeneities of the ice cover, which is a conservative process that redistributes energy in all directions,
and (ii) dissipative processes which are related to various sources,
e.g. friction due to the presence of sea ice, inelastic collisions and breakup of ice floes.
The relative importance of scattering and dissipation is still unclear,
and uncertainties still exist about the actual mechanisms for wave dissipation in sea ice.
This has led to a surge of research activity on this topic in recent years.
Studies have suggested that dissipative processes are dominant in frazil and pancake ice fields \cite{ddmbw15,nm99},
while scattering seems to be the main mechanism for wave attenuation in broken floe fields \cite{km08,wsgcm88}.
Even for a denser ice field, the problem remains complex:
e.g. Ardhuin et al. \cite{asdw16} found that dissipation dominates over scattering for long swells in the Arctic ice pack.

With a view to describing wave attenuation in the MIZ, two different approaches have been pursued
based on linear theory: (i) discrete-floe models where individual floes with possibly distinct characteristics are resolved
assuming an idealized geometry \cite{bs12,msb16,wbsdb13}, and (ii) continuum models where the heterogeneous ice field is viewed 
as a uniform material with effective rheological properties including viscosity or viscoelasticity \cite{dd02,k98,ws10}.
In case (i), the analysis focuses on wave scattering and typically requires solving a boundary value problem 
with multiple regions in the horizontal hyperplane.
By contrast, case (ii) enables the derivation of an exact algebraic expression for the dispersion relation governing traveling plane waves
in the effective medium. Wave attenuation (possibly from scattering and dissipation combined) is encoded in the complex roots
of this dispersion relation, and various physical effects are controlled by constant parameters.
Recent reviews on this theoretical work can be found in \cite{s19,s20}.

Models of type (i) have been applied to various floe configurations and have reached a high degree of sophistication.
There is now a consensus that the process of wave scattering in sea ice is well understood,
and associated parameterizations have been tested for operational wave forecasting \cite{db13,ph96,wbsdb13b}.
It is however not the case for dissipative processes and, partly for this reason, there has been an increasing effort in recent years
at developing and calibrating models of type (ii) \cite{cheng17,desanti18,ddmbw15,rogers16}.
In this framework, details of the attenuation processes are not accounted for and it lies on the calibration
to ensure that the effective parameters are assigned suitable empirical values for practical applications.

While continuum models have been employed for some time now, based mostly on thin-plate theory,
to describe wave propagation in pack ice \cite{fs94,lm88}, 
their extension to the setting of a more compliant or fragmented ice cover for application to the MIZ is more recent \cite{zsc15}.
Earlier versions include the two-layer viscous model of Keller \cite{k98} which treats the ice cover as
a viscous layer lying on top of an ideal fluid (the ocean).
The viscous layer is meant to represent a suspension of ice particles in water.
Interaction among these particles and the associated friction leads to wave energy dissipation, which is modeled as a viscous effect.
Good agreement has been found in comparison to laboratory data on wave attenuation in grease ice \cite{nm97,nm99}.
Keller's model was extended by de Carolis and Desiderio \cite{dd02} to allow for a viscous fluid in the lower layer as well.
Validation was provided to some extent against laboratory and field measurements.
Wang and Shen \cite{ws10} refined Keller's model by adding elasticity to the upper layer 
as this property may be of relevance to broken floe fields.
Their viscoelastic model has been tested against laboratory experiments under various ice conditions,
and has been calibrated and used in parameterization of wave hindcasts for the Arctic and Antarctic.
Building upon this idea, Zhao and Shen \cite{zs18} developed a three-layer version which features a turbulent boundary layer 
between the viscoelastic ice cover and the inviscid ocean.
Dissipation due to turbulence in the middle layer is associated with some eddy viscosity.

In the spirit of this continuum approach, Chen et al. \cite{cgg19} recently proposed a more elaborate two-layer model
where the ice cover is viewed as a homogeneous isotropic poroelastic material according to Biot's theory \cite{cgg18}.
More specifically, the heterogeneous ice field is described as a mixed layer with a solid phase and a fluid phase
as the two limiting configurations. Each phase is assumed to be slightly compressible.
Dissipative effects are included via two different mechanisms: viscosity within each phase of the ice layer,
and friction caused by the relative motion between its fluid and solid constituents.
A parameter of interest in this model is the ice porosity which may serve to provide a measure of ice concentration.
Despite the complicated nature of this formulation, an exact linear dispersion relation can be derived 
and numerical estimates of physically relevant solutions can be found using relatively simple selection criteria.
Preliminary tests were conducted in \cite{cgg19} to verify consistency with predictions from simpler models
(e.g. open water, mass loading, purely elastic) in their respective limits \cite{crl17,xg09}.
A more detailed review of this dispersion relation together with those from other viscoelastic representations
will be presented in the next section.

The main goal of this paper is to further assess the porous viscoelastic model of Chen et al. \cite{cgg19}
by testing it against both laboratory experiments and field observations of wave attenuation in sea ice.
These are taken from the literature, and allow us to probe a wide range of ice conditions and wave frequencies.
This is accomplished by fitting the theoretical predictions to data on attenuation rate 
via error minimization over a set of rheological parameters.
As a result, numerical estimates for both the attenuation rate and the set of effective parameters are obtained from the fitting process.
This model's performance is also checked by comparing it to other existing viscoelastic formulations
under the same various conditions.
The purpose of such a comparison is two-fold. First, it helps examine in detail the parametric dependence 
in viscoelastic theories, and  the extent to which common rheological parameters may differ in their range of values.
Indeed, this difference may be of several orders of magnitude for such effective parameters.
Second, it helps validate our data fitting method as we can check with previous independent calibration results from the literature.
Given the relatively large parameter space in this poroelastic setting, a sensitivity analysis is also performed 
to gauge the individual contributions of rheological parameters to the fitting process.
A notable finding from our study is that Chen et al.'s model can reproduce to some degree the roll-over of attenuation rate 
as observed in field measurements from the Arctic MIZ.
This intriguing phenomenon has generally eluded linear scattering or viscoelastic models
and, while various possible causes have been suggested, it is still not well understood \cite{lkdwgs17,thmkk21}.

The remainder of this paper is organized as follows.
Section 2 recalls the linear dispersion relation obtained from the porous viscoelastic model of Chen et al. \cite{cgg19}
and describes the data fitting procedure.
Other existing viscoelastic formulations are also briefly reviewed.
Section 3 presents the corresponding fits to data on attenuation rate from a selection of laboratory experiments and field observations.
Section 4 discusses the estimation of shear modulus and kinematic viscosity, 
and compares results among three different viscoelastic models.
Section 5 shows sensitivity tests on a set of rheological parameters that are relevant to the poroelastic system.
Finally, concluding remarks are provided in Section 6.

\section{Theoretical models}

The dispersion relations reviewed in this section are derived from continuum models for linear traveling waves
in the two-dimensional case (one horizontal direction and one vertical direction).

The dispersion relation associated with the porous viscoelastic model proposed in \cite{cgg19} 
(hereafter referred to as CGG) is given by
\begin{equation} \label{poro-disp}
\omega^{2} = \left( \frac{T_1 + g \, T_2}{T_3} \right) D_{4} \tanh(D_{4} H) \,,
\end{equation}
with
\[
D_4 = \sqrt{\kappa^2 - \frac{\omega^2}{c_f^2}} \,,
\]
where $g$ is the acceleration due to gravity, $H$ is the water depth and $c_f$ is the speed of sound in water.
The reader is directed to \cite{cgg19} for a detailed derivation of this model and to the Appendix 
where the expressions of coefficients $T_1$, $T_2$, $T_3$ are recalled for convenience.
These coefficients are functions of various wave parameters and rheological parameters.
Wave parameters include the angular frequency $\omega$ and complex mode $\kappa = k + {\rm i} \, q$ 
where $k$ is the wavenumber and $q$ is the attenuation rate.
Rheological parameters include the water density $\rho_f$ as well as the ice density $\rho_s$, porosity $\beta$, 
shear modulus $\mu$, Poisson's ratio $\nu$, kinematic viscosity $\eta$ and thickness $h$.
Ice porosity is represented by a dimensionless parameter whose range is $0 \le \beta \le 1$,
with the limiting values $\beta = 0$ (solid phase) and $\beta = 1$ (fluid phase) corresponding to pack ice and near-open water, respectively.
This parameter may be related to ice concentration $C$ via the relation $\beta = 1 - C$,
namely $\beta$ is the complement of $C$.
It should be pointed out that, in this continuum framework, the elasticity, porosity and viscosity parameters
do not necessarily correspond to intrinsic properties of sea ice but rather they are meant to represent 
effective properties of the heterogeneous ice field under consideration,
similar to e.g. homogenization modeling of wave propagation in complex media \cite{cgs09,dcdgs08}.
These parameters may thus vary over a wider range than the typical values for sea ice.
In view are potential applications to large-scale wave forecasting in the MIZ where various types of sea ice coexist.
Whenever the information is available, $\beta$ and $h$ will be assigned values corresponding to 
mean ice concentration and mean thickness of the ice cover
(e.g. field studies often report an estimate of the fraction of ice-covered surface that may be used for $C$).
Aside from bulk viscosity which is typically regulated by $\eta$, this model also describes friction due to
the relative motion between fluid and solid parts of the ice field. 
In the equations, the coefficient controlling this mechanism is defined by
\begin{equation} \label{friction}
b = \frac{8 \rho_s \eta \beta}{a^2} \,,
\end{equation}
where $a$ denotes the fluid pore size in the porous medium. 
From the viewpoint of effective medium theory for wave propagation in the MIZ,
this parameter $a$ may be related to some characteristic horizontal size of open-water areas in the fragmented ice cover.

In the next section, predictions from \eqref{poro-disp} will be tested against a selection of laboratory experiments
and field observations. For each set of experimental data, comparison with other existing models will be provided as well.
These include recent viscoelastic models by Wang and Shen \cite{ws10} and Mosig et al. \cite{mms15},
which we find convenient to present in detail below because they share common rheological parameters,
and this will be of relevance to the subsequent discussion.
These two models are simpler than the present one in the sense that they do not take into account ice porosity,
accordingly their dispersion relations are simpler.
The dispersion relation resulting from Wang and Shen's model \cite{ws10} (hereafter referred to as WS) can be written as
\begin{equation} \label{WS}
\omega^2 = \left( 1 + \frac{\rho_s N_3}{\rho_f N_4} \right)  g {\kappa} \tanh({\kappa} H) \,,
\end{equation}
where
\begin{eqnarray*}
N_4 & = & g {\kappa} \big[ 4 {\kappa}^3 N_1 \eta_c^2 \sinh({\kappa} h) \cosh(N_1 h) + N_2^2 \cosh({\kappa} h) \cosh(N_1 h) \\
& & - g {\kappa} \sinh({\kappa} h) \sinh(N_1 h) \big] \,, \\
N_3 & = & (g^2 {\kappa}^2 - N_2^4 - 16 {\kappa}^6 N_1^2 \eta_c^4) \sinh({\kappa} h) \sinh(N_1 h) \\
& & - 8 {\kappa}^3 N_1 \eta_c^2 N_2^2 \big[ \cosh({\kappa} h) \cosh(N_1 h) - 1 \big] \,, 
\end{eqnarray*}
and
\[
N_2 = \omega + 2 {\rm i} \, \eta_c {\kappa}^2 \,, \quad
N_1 = \sqrt{{\kappa}^2 - {\rm i} \frac{\omega}{\eta_c}} \,, \quad
\eta_c = \eta + {\rm i} \frac{\mu}{\rho_{s} \omega} \,.
\]
On the other hand, the dispersion relation produced by Mosig et al.'s model \cite{mms15} (hereafter referred to as EFS) takes the form
\begin{equation} \label{EFS}
\frac{\omega^{2} }{g - \frac{\rho_s \omega^{2} h}{\rho_f} + \frac{\mu_c h^3 \kappa^{4}}{6 (1 - \nu) \rho_f}} = \kappa \tanh(\kappa H) \,,
\end{equation}
where 
\begin{equation} \label{mu}
\mu_c = -{\rm i} \, \rho_s \omega \eta_c = \mu -{\rm i} \, \omega \rho_{s} \eta \,. 
\end{equation}
Note that the EFS model also makes use of the thin-plate approximation and thus is significantly simpler than 
both CGG and WS models which instead consider the ice cover as a distinct layer with an actual thickness.
Preliminary comparison between these three models can be found in \cite{cgg19}.

For a given value of $\omega$ and other parameter values, the dispersion relation is solved numerically for $\kappa$ 
using the root-finding routine \textit{fsolve} in Matlab.
More specifically, because $\kappa$ is complex, Eq. \eqref{poro-disp} is split up into its real and imaginary parts.
This leads to a system of two independent equations that are solved simultaneously for the two unknowns $k$ and $q$.
The \textit{fsolve} algorithm is essentially a quasi-Newton method with a numerical approximation of the Jacobian matrix.
We have successfully used this Matlab routine in previous work \cite{g06,gp14} to compute traveling wave solutions
of nonlinear partial differential equations.
Considering that multiple roots for $k$ and $q$ may exist here \cite{mms15,zcs17}, we apply the selection criteria proposed in \cite{ws10}
to find a dominant pair $(k,q)$ associated with a physically relevant solution.
We choose the converged values $(k,q) \in \mathbb{R}_+^2$ such that $k$ is closest to the open-water wavenumber $k_0$ 
which solves 
\[
\omega^2 = g k \tanh(k H) \,,
\]
and $q$ is the lowest attenuation rate possible.
Accordingly, we run the root finder \textit{fsolve} for a range of initial guesses around $(k,q) = (k_0,0)$ 
and select the converged values for which the error $| \kappa - (k_0 + {\rm i} \, 0)|$ is minimum among all the roots found.
In doing so, we were able to get acceptable solutions in all the cases we considered.

For the following tests, we prescribe a number of physical parameters such as 
$g = 9.81$, $\rho_s = 917$, $\rho_f = 1025$ and $c_f = 1449$ (in SI units), 
and fit the model predictions to the experimental data by optimizing with respect to other parameters.
The water depth $H$ is also specified, since this is either known information from laboratory experiments
or a representative value may be used for a specific oceanic region when comparing to field observations.
Furthermore, because the range of Poisson's ratio is typically small ($0 < \nu < 1/2$), 
we set it to be $\nu = 0.4$, after checking that it does indeed not play a major role (see Section 5).
This reduction in the parameter space helps simplify the analysis, which is especially relevant for the CGG model
considering that it involves many physical parameters.

We will focus our attention on the subset $(\beta, \mu, \eta)$ when fitting the model predictions to the experimental data.
Throughout this study, we will only consider data on the attenuation rate, a reason being that data on the wavenumber 
were not reported by field observations and our focus here is on the attenuation process as in \cite{km08,mms15,ph96,srcj19}.
Our fitting procedure is basically a direct search approach. For a given triplet $(\beta, \mu, \eta)$ and a range of values of $\omega$, 
we apply the above-mentioned root-finding scheme to find a set of pairs $(k,q)$. We repeatedly run this procedure 
over a specified region of parameter space $(\beta, \mu, \eta)$, and select the set $\{ q_j \}$ for which the $L^2$ error
\begin{equation} \label{error}
E = \sum_{j=1}^n (q_j - \widehat q_j)^2 \,,
\end{equation}
between numerical estimates $\{ q_j \}$ and experimental data $\{ \widehat q_j \}$,
is minimum among all the solutions calculated.
The best fit so obtained returns a set $\{ q_j \}$ for the attenuation rate,
as well as a triplet $(\beta, \mu, \eta)$ for these rheological parameters.
For the problem at hand, we use a straightforward definition \eqref{error} of the error as in \cite{desanti18},
which is readily applicable to all the cases explored and which allows for a direct comparison among the various models involved.

It turns out that information on ice concentration was also reported in some of these studies
(which we used to determine the ice porosity), and thus only the pair of parameters $(\mu,\eta)$ is to be found from the data fitting.
In this process, the ranges of values for $(\mu,\eta)$ and their resolutions are chosen in a heuristic manner based on extensive trials,
considering previous work \cite{mms15,nm99,zs15} and our own experience \cite{cgg19,cgg18}.
The larger the region of parameter space to be covered, the higher the computational cost.
Typically, we conduct a preliminary search over an extended rough region of parameter space 
and then refine the search over smaller better-resolved sectors.
Aside from testing the performance of the CGG model, such an analysis also helps calibrate it 
by estimating rheological parameters for potential applications in realistic conditions.
As discussed below, while the CGG, EFS and WS models share common physical parameters such as $\mu$ and $\eta$,
their respective numerical values according to the data fitting may differ significantly.
Note that we use the same procedure to obtain fitting curves from the EFS and WS models.

Although it is somewhat different in character from the CGG, EFS and WS models, we will also include a comparison with
the two-layer viscous model recently proposed by Sutherland et al. \cite{srcj19} (hereafter referred to as SRCJ),
which estimates wave attenuation in sea ice by
\begin{equation} \label{viscous}
q = \frac{1}{2} \Delta_0 \epsilon h k_0^2 \,.
\end{equation}
This formula is partly heuristic because it was derived based on scaling arguments and dimensional analysis.
It is nonetheless appealing due to its stunning simplicity and has been shown to produce satisfactory results
in comparison with experimental data.
Considering that viscous models have been successful at describing wave attenuation in such ice covers
as grease ice, predictions from \eqref{viscous} may serve as a suitable independent reference,
especially for the tests involving laboratory experiments.
The coefficients $\Delta_0$ and $\epsilon$ are dimensionless empirical parameters whose range is 
$0 < \Delta_0 < 1$, $0 < \epsilon < 1$.
The same $L^2$ error \eqref{error} is used to optimize \eqref{viscous} with respect to the pair $(\Delta_0,\epsilon)$ 
when fitting to measurements.
The parameter $\Delta_0$ is a measure of the relative motion between ice and water at the bottom boundary of the ice layer. 
To first approximation, $\Delta_0 \simeq 1$ corresponding to a no-slip boundary condition.
The parameter $\epsilon$ is thought to be a function of ice porosity, exhibiting a similar behavior.
In particular, the limit $\epsilon \rightarrow 0$ is analogous to $\beta \rightarrow 0$ (pack ice)
for which, according to \eqref{viscous}, there should be no wave dissipation within the ice layer.
Sutherland et al. \cite{srcj19} pointed out that such a parameterization is consistent with observations of wave attenuation
through the MIZ, being several orders of magnitude greater in frazil and pancake ice than in a broken floe field \cite{ddmbw15}.
This is somewhat counterintuitive considering that the latter represents a more rigid ice cover than the former.
Note however that the dominant mechanism for wave attenuation in broken floe fields is believed to be scattering
and is of different nature from the viscous-type dissipation taking place in pancake ice fields.
Preliminary tests of the CGG model in \cite{cgg19} are also consistent with these observations
and indicate a tendency for $q$ to decrease as $\beta \rightarrow 0$ over all frequencies.

\section{Comparison with experimental data}

In this section, we test the CGG model against three different sets of laboratory experiments and two different sets of field observations.
Altogether, these span a wide range of wave frequencies and ice-cover types.
In each case, the CGG model is tested by best fitting its predictions to experimental data on the attenuation rate,
based on the numerical scheme described earlier.
Predictions by other existing models are also shown for comparison and numerical values of their rheological parameters
as determined by the data fitting are discussed.

\subsection{Laboratory experiments of Newyear and Martin (1997)}

Newyear and Martin \cite{nm97} conducted laboratory experiments of wave propagation and attenuation in grease ice.
This study was among the first to measure wave attenuation by floating ice in a controlled laboratory environment.
The wave tank was a flat-bottomed rectangular box, $3.5$ m long, $1$ m wide and $1$ m deep.
They reported two sets of measurements for two different ice thicknesses $h = 11.3$ cm (Test 1) and $h = 14.6$ cm (Test 2).
In both cases, the water depth was set to $H = 0.5$ m and the ice concentration was estimated to be $C = 0.53$.

\begin{figure}[t!]
\centering
\begin{subfigure}[h]{0.52\textwidth}
\includegraphics[width=\textwidth]{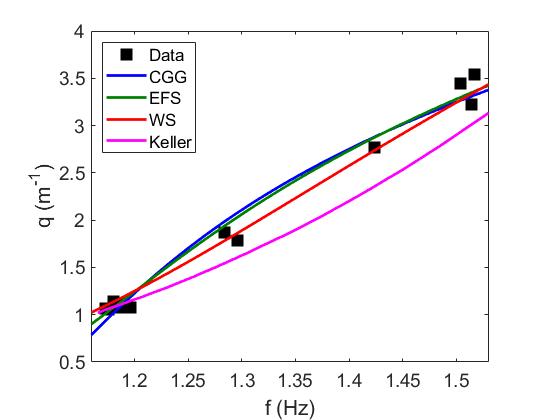}
\caption{$h=11.3$ cm}
\label{fig:Newyear_table_1}
\end{subfigure}\hspace*{\fill}
\begin{subfigure}[h]{0.52\textwidth}
\includegraphics[width=\textwidth]{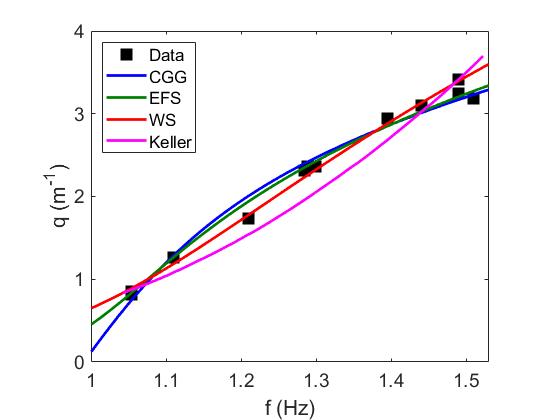}
\caption{$h=14.6$ cm}
\label{fig:Newyear_table_2}
\end{subfigure}
\caption{Comparison of attenuation rate vs. frequency between model predictions and laboratory data for grease ice from Newyear and Martin \cite{nm97}.
Predictions from the CGG, EFS, WS and Keller's models are shown.
Laboratory data for (a) $h = 11.3$ cm (test 1), (b) $h = 14.6$ cm (test 2) are presented.}
\label{fig:Newyear}
\end{figure}

Figure \ref{fig:Newyear} shows best fits of the CGG, EFS and WS models to Newyear and Martin's measurements
of attenuation rate $q$ (from their Tables 1 and 2) as functions of wave frequency $f = 2\pi \omega$.
The value $\beta = 1 - C = 0.47$ for ice porosity is used in the CGG model.
Newyear and Martin \cite{nm99} provided a comparison of their laboratory data with Keller's two-layer viscous model \cite{k98},
whose predictions are also shown in Fig. \ref{fig:Newyear} (these were extracted from figures in their paper).
Keller's model is basically a counterpart of WS model without elasticity.
The agreement between the CGG model and the experiments is fairly good in both cases.
We see that the CGG and EFS curves are particularly close together. 
They appear to be concave while Keller's curve appears to be convex, 
which is characteristic of a purely viscous (i.e. diffusive-type) mechanism \cite{lhv91,srcj19}.
The CGG concavity is especially pronounced for $h = 14.6$ cm (Fig. \ref{fig:Newyear}b),
which leads to a close fit at high frequencies where the increase of $q$ seems to slow down.
As for the WS curve, it behaves more linearly with respect to $f$ and lies between these two opposite trends.
Note that it is not clear whether the actual trend is concave or convex due to measurement errors
and the limited number of data points.

\subsection{Laboratory experiments of Wang and Shen (2010)}

This study was conducted as part of the RECARO (REduced ice Cover in the ARctic Ocean) project
in the Arctic Environmental Test Basin at Hamburg Ship Model Basin (HSVA), Germany.
The wave basin was roughly $19$ m long, $6$ m wide and $1.5$ deep, and was separated equally
lengthwise into two $3$ m wide flumes (referred to as Tank 2 and Tank 3).
Experiments were performed in these two flumes to measure wave propagation and attenuation in a grease-pancake ice mixture \cite{ws10b}.
The ice thickness was not uniform in these experiments, so we use the mean values 
$h = 9.0$ cm and $8.9$ cm for Tank 2 and 3 respectively.
The water depth was $H = 0.85$ m but no information was provided on the ice concentration.

\begin{figure}[t!]
\centering
\begin{subfigure}[h]{0.52\textwidth}
\includegraphics[width=\textwidth]{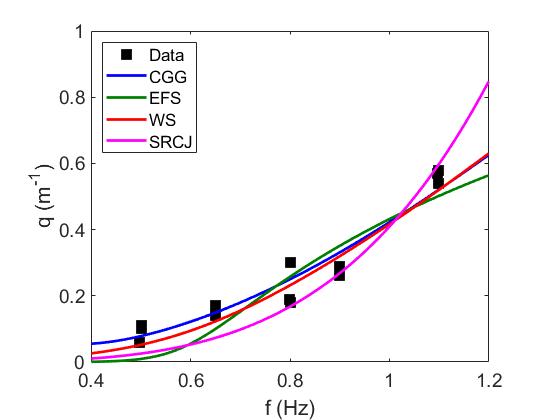}
\caption{$h=9.0$ cm}
\label{fig:WangShen tank 2}
\end{subfigure}\hspace*{\fill}
\begin{subfigure}[h]{0.52\textwidth}
\includegraphics[width=\textwidth]{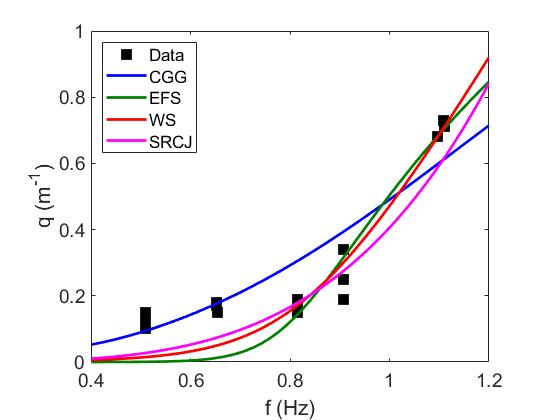}
\caption{$h=8.9$ cm}
\label{fig:WangShen tank 3}
\end{subfigure}
\caption{Comparison of attenuation rate vs. frequency between model predictions and laboratory data for a grease-pancake ice mixture from Wang and Shen \cite{ws10b}.
Predictions from the CGG, EFS, SRCJ and WS models are shown.
Laboratory data for (a) $h = 9.0$ cm (tank 2), (b) $h = 8.9$ cm (tank 3) are presented.}
\label{fig:WangShen}
\end{figure}

Comparison of these experimental data (from Tables 1 and 2 in \cite{ws10b}) 
with the CGG, EFS, SRCJ and WS predictions is given in Fig. \ref{fig:WangShen}.
The general trend appears to be convex for most of these models.
The EFS fit falls down very quickly as $f \rightarrow 0$ but seems to develop an inflection 
(from convex to concave) while rising up at high frequencies.
The CGG fit looks satisfactory overall. It does not quite capture the high convexity around $f = 0.9$ Hz
(in particular for Tank 3) but it does not fall down as quickly as the other curves at lower frequencies, 
which is consistent with the asymptotic behavior suggested by the experimental results in that limit.
A similar observation was made by Wang and Shen \cite{ws10b} who found that a grease-pancake ice layer
appears to be more dissipative (producing a higher attenuation rate) in the low-fequency range than predicted by
Keller's viscous model, and concluded that such a mixed ice layer may be rheologically quite different from
a pure grease ice layer (for which a viscous model usually works well).

The CGG fit estimates the ice porosity to be $\beta = 0.16$ and $0.15$ for Tank 2 and 3 respectively.
These low values of $\beta$ mimic a configuration where the ice cover is relatively compact,
which is compatible with the presence of pancake ice, and as expected they are lower than the value
$\beta = 0.47$ deduced from Newyear and Martin's measurements for grease ice.
The associated pore sizes are given by $a = 1.2$ cm and $2.4$ cm for Tank 2 and 3.
Recall that the pores represent the fluid part of the porous ice cover in the continuum formulation of the CGG model.
For low $\beta$, we may thus assume that constitutive elements of the solid part would have a typical size
on the same order of magnitude as or larger than $a$,
which is consistent with the pancake diameter ranging from about $\ell = 1$ cm to $40$ cm 
as observed in Wang and Shen's experiments.

\subsection{Laboratory experiments of Zhao and Shen (2015)}

Zhao and Shen \cite{zs15} followed up with additional experiments at the HSVA in 2013.
Three sets of measurements were performed in Tank 3 (as defined in the previous section)
for three different types of ice cover:
a frazil/pancake ice mixture (with thickness $h = 2.5$ cm), pancake ice ($h = 4.0$ cm)
and a broken floe field ($h = 7.0$ cm).
These three cases are referred to as Test 1, 2 and 3 respectively.
The water depth was about $H = 0.94$ m and again, although values for the diameter of a typical pancake/floe
were reported in \cite{zs15}, no information was given on the mean ice concentration for the generated ice fields.

\begin{figure}[t!]
\centering
\begin{subfigure}[b]{0.55\textwidth}
\includegraphics[width=1\linewidth]{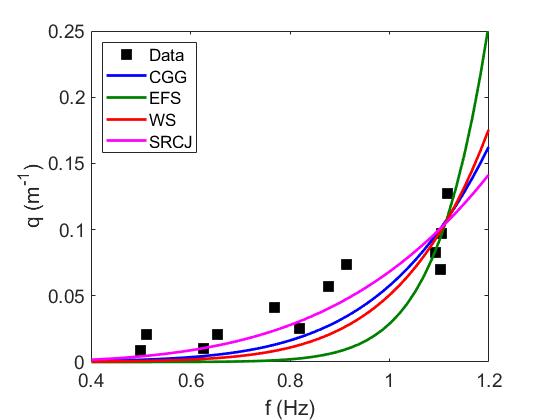}
\caption{$h=2.5$ cm}
\label{fig:test1} 
\end{subfigure}
\begin{subfigure}[b]{0.55\textwidth}
\includegraphics[width=1\linewidth]{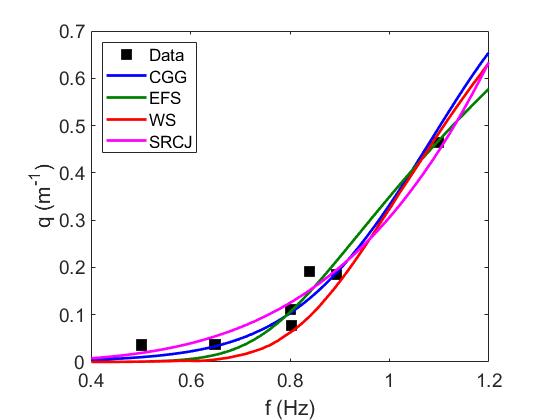}
\caption{$h=4.0$ cm}
\label{fig:test2}
\end{subfigure}
\begin{subfigure}[b]{0.55\textwidth}
\includegraphics[width=1\linewidth]{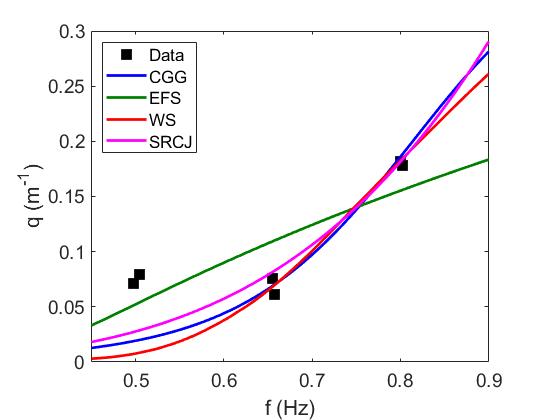}
\caption{$h=7.0$ cm}
\label{fig:test3}
\end{subfigure}
\caption{Comparison of attenuation rate vs. frequency between model predictions and laboratory data from Zhao and Shen \cite{zs15}.
Predictions from the CGG, EFS, SRCJ and WS models are shown.
Laboratory data for (a) $h = 2.5$ cm (test 1, frazil/pancake ice), (b) $h = 4.0$ cm (test 2, pancake ice), (c) $h = 7.0$ cm (test3, fragmented ice) are presented.}
\label{fig:ZhaoShen}
\end{figure}

Overall, the models compare well with these experiments (from Table 3 in \cite{zs15}), as indicated in Fig. \ref{fig:ZhaoShen}.
Their fits are especially good for Tests 1 and 2, and are reminiscent of the previous results (Fig. \ref{fig:WangShen})
with Wang and Shen's experiments for a grease-pancake ice mixture.
The SRCJ model is found to perform quite well over the entire range of frequencies considered, 
even at low frequencies where, despite tending to zero, its fitting curve is closest to the data points.
This contrasts with a previous observation regarding the comparison to Wang and Shen's experiments,
and may be explained by the fact that the ice thicknesses for Tests 1 and 2 are significantly smaller than
those specified in \cite{ws10b}.
The agreement is less convincing for Test 3, partly because there are fewer data points available.
These suggest a convex dependence of $q$ on $f$, which is captured to some extent by the CGG, SRCJ and WS models.
The laboratory measurements however yield much higher attenuation rates at low frequencies than what these models predict, 
indicating a tendency for $q$ to saturate or even increase back as $f$ decreases.
The EFS curve looks quite different from the other curves, exhibiting a slightly concave profile.
It is worth pointing out that all these models underestimate the attenuation rate at low frequencies for all three experiments.

From the CGG fit, we find $\beta = 0.01$, $0.07$ and $0.06$ for Test 1, 2 and 3 respectively.
The corresponding pore sizes are $a = 2.0$ cm, $4.8$ cm and $8.5$ cm.
Note the particularly low level of ice porosity that is obtained for Test 1.
An interpretation for this case is that the ice thickness is so small and consequently the wave attenuation is so weak
(as indicated by the low attenuation rates in Fig. \ref{fig:ZhaoShen}a) that the CGG model views it as equivalent to 
a configuration with pack ice (corresponding to the limit $\beta \rightarrow 0$).
This is consistent with the parameter values $\epsilon = 0.34$, $0.94$ and $0.85$ estimated from the SRCJ fits
for Test 1, 2 and 3 as reported by Sutherland et al. \cite{srcj19}.
These authors also tested their viscous model against the laboratory experiments of Newyear and Martin \cite{nm97},
Wang and Shen \cite{ws10b} and Zhao and Shen \cite{zs15}, and found the parameter $\epsilon$ to be smallest
for Test 1 of Zhao and Shen among all the cases considered.
We obtain a similar result here, with $\beta$ being smallest for that particular experiment.
Moreover, the set of $\beta = \{ 0.01, 0.07, 0.06 \}$ seems to follow a pattern of variation
similar to that for the set of $\epsilon = \{ 0.34, 0.94, 0.85 \}$.
Again, for such low levels of ice porosity as given by the CGG fit, the associated pore sizes 
$a = \{ 2.0, 4.8, 8.5 \}$ cm may be deemed compatible with the typical ice diameters
$\ell = \{ 3, 5, 20 \}$ cm observed in \cite{zs15} for Test 1, 2, 3.

We point out in passing that Sutherland et al. \cite{srcj19} set $\Delta_0 = 1$ and determined $\epsilon$
by a linear least-squares fitting method.
In the present study, we fit \eqref{viscous} to the experimental data by minimizing \eqref{error}
with respect to both $\Delta_0$ and $\epsilon$, based on the approach described in Section 2.
This produces values of $\Delta_0$ near $1$ and values of $\epsilon$ that are very close to those reported in \cite{srcj19},
which may serve as evidence for the effectiveness of our fitting method.
From the SRCJ fit shown in Fig. \ref{fig:ZhaoShen}, we find 
$(\Delta_0,\epsilon) = (0.96,0.34)$, $(0.97,0.97)$ and $(0.94,0.82)$ for Test 1, 2 and 3.

Zhao and Shen \cite{zs15} also used their laboratory data to test the WS model and estimate such parameters as
the shear modulus and kinematic viscosity. We will refer to their results as part of the discussion in Section 4.

\subsection{Field observations of Wadhams et al. (1988)}

During field operations in the Greenland and Bering Seas in the late 1970s and early 1980s,
the Scott Polar Research Institute \cite{wsgcm88} carried out a series of experiments 
where wave attenuation was measured along a line of stations running from the open sea deep into an ice field.
Large broken floes are a prominent feature of the ice field in this case.
At each station, a wave buoy was inserted between floes to measure the local wave spectrum.
A mean ice thickness was determined by coring at each of the experimental floes
along the major axis of the incoming wave spectrum.
Floe size distributions were derived from overlapping vertical photography from a helicopter.
Among the measurements reported in \cite{wsgcm88} (see their Table 2), 
we will use those from the Greenland Sea in 1979 and from the Bering Sea in 1983.
Other data sets (e.g. 1978 Greenland Sea and 1979 Bering Sea) were deemed not suitable
due to possibly larger experimental error or unwanted physical effects such as 
wave reflection/absorption from the fjords, as mentioned in \cite{km08}.
We will take this opportunity to compare with results of Kohout and Meylan \cite{km08}
(hereafter referred to as KM) who also tested their scattering model against these field observations.

An intriguing feature of the 1979 Greenland Sea and 1983 Bering Sea measurements is that 
they show a roll-over of attenuation rate as a function of wave period (or wave frequency), in lieu of a monotonic behavior.
This roll-over occurs at short periods (or high frequencies) in the range considered.
Continuum viscoelastic models or discrete scattering models have usually been unable to predict this phenomenon.
Possible explanations that have been suggested include wind forcing, 
nonlinear wave interactions or instrument noise \cite{lkdwgs17,ph96,thmkk21,wsgcm88}.
An exception that we are aware of in the context of linear theory is the three-layer viscoelastic model with eddy viscosity
as recently proposed by Zhao and Shen \cite{zs18}.
Their numerical results show a roll-over that accentuates as the thickness of the turbulent boundary layer 
(located between the viscoelastic ice layer and the inviscid water layer) increases.
However, no comparison with field data featuring the roll-over was presented in that study.
A similar phenomenon was observed by Liu et al. \cite{lhv91}
based on a linear model for a thin elastic plate with eddy viscosity \cite{lm88}.
These authors derived a temporal rate of wave attenuation and converted it to a spatial rate
by dividing it by the group velocity.
As noted in \cite{lkdwgs17}, this temporal rate is a monotonic function of frequency 
and so the fact that the spatial rate exhibits roll-over is likely due to the group velocity 
being non-monotonic and reaching a minimum at some frequency \cite{gp12}.
Therefore, it is not clear from this result whether the roll-over effect is an intrinsic feature of the thin-plate viscoelastic model
or is simply an artifact of the observation procedure.

\subsubsection{Greenland Sea, 10 September 1979}

During this experiment, the ice cover was sparse and the floes were generally large.
The ice concentration was estimated to be $C = 0.17$ from photograph analysis.
Because ice thicknesses could not be determined on that day and were not reported, we choose $h = 3.1$ m
following Kohout and Meylan \cite{km08} who suggest using the floe thickness from the 1978 data,
which was based on 14 measurements through smooth areas.
We set $H = 1500$ m (average depth of the Greenland Sea) and,
for the CGG model, we assign the value $\beta = 1 - C = 0.83$ to ice porosity.

\begin{figure}[t!]
\centering
\includegraphics[width=\linewidth]{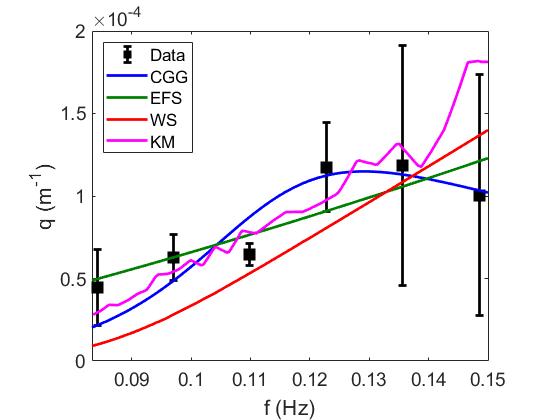}
\caption{Comparison of attenuation rate vs. frequency between model predictions and data for a broken floe field from the Greenland Sea 10 September 1979 experiment in Wadhams et al. \cite{wsgcm88}.
Predictions from the CGG, EFS, KM and WS models are shown.
Results for $h = 3.1$ m and $H = 1500$ m are presented.}
\label{fig:Greenland}
\end{figure}

Note that the field measurements under consideration are on the wave spectrum,
which is proportional to the square of the wave amplitude.
Accordingly, we halve the corresponding decay rates when comparing to
theoretical predictions for the wave amplitude.
We see in Fig. \ref{fig:Greenland} that the CGG model fits the field observations well,
despite the small number of data points available.
Among all the models at play, it is the only one that is able to reproduce some roll-over of $q$, with a peak near $ f = 0.13$ Hz.
In fairness, we should mention that experimental errors are more appreciable at this end of the spectrum.
The CGG model also captures the stronger decay of attenuation rate at lower frequencies,
although its predictions of $q$ tend to be even lower than the measured values in the limit $f \rightarrow 0$.
By contrast, the EFS and WS curves are monotonically increasing with $f$, 
almost linearly over the range of frequencies considered.
The SRCJ fit is also found to be monotonically increasing with frequency and is not plotted in this figure.

Instead, we show the KM fit which is extracted from Fig. 8 in \cite{km08}
(with appropriate rescaling to convert the dimensionless energy rates per floe number in that figure 
to dimensional rates of spatial attenuation for the wave amplitude).
We point out that the KM model is a scattering model and is of different nature from
the continuum formulation that is highlighted in the present study.
It is thus not further discussed here and the reader is directed to \cite{km08} for more detail.
Because scattering is believed to be the dominant mechanism for wave attenuation in broken floe fields,
the KM model serves as a suitable independent reference for the comparison with field observations from \cite{wsgcm88}.
The KM curve appears to be rougher than the other theoretical curves,
as it represents the average of 100 simulations with different random realizations of the floe size distribution.
We can nonetheless discern a general trend that is monotonically increasing with frequency,
and is approximately linear with a slope close to that of the WS curve.

The CGG fit presented in Fig. \ref{fig:Greenland} returns a pore size $a = 14.6$ m.
While it is difficult to give a physical interpretation for this parameter from the viewpoint of effective medium theory,
we may associate it to a characteristic horizontal size of open-water areas in the context of an extensive broken floe field.
It is reassuring that we find a value of $a$ which is significantly larger than those obtained for 
the (smaller-scale) laboratory experiments of Wang and Shen \cite{ws10b} and Zhao and Shen \cite{zs15}.
Moreover, although pore size in the CGG model does not signify floe size as mentioned earlier
(and these two parameters are not necessarily correlated), we deem it consistent that the estimated value $a = 14.6$ m
is somewhat comparable in order of magnitude to the typical floe size ($\ell = 50$--$80$ m) 
observed on this expedition, as reported in \cite{km08}.

\subsection{Bering Sea, 7 February 1983}

This experiment was carried out as part of the MIZEX West study in 1983.
Following Perrie and Hu \cite{ph96} and Kohout and Meylan \cite{km08},
we take representative values for the ice concentration and thickness in this case 
to be $C = 0.72$ and $h = 1.5$ m, respectively.
The ice cover was thus less fragmented than in the previous environment.
We select $H = 1500$ m for the average depth of the Bering Sea and prescribe $\beta = 1 - C = 0.28$ in the CGG model.

\begin{figure}[t!]
\centering
\includegraphics[width=\linewidth]{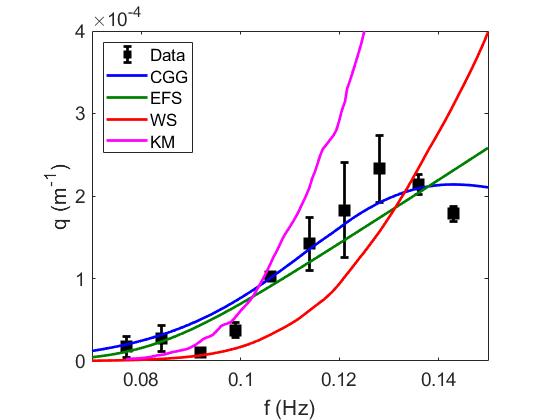} 
\caption{Comparison of attenuation rate vs. frequency between model predictions and data for a broken floe field from the Bering Sea 7 February 1983 experiment in Wadhams et al. \cite{wsgcm88}.
Predictions from the CGG, EFS, KM and WS models are shown.
Results for $h = 1.5$ m and $H = 1500$ m are presented.}
\label{fig:Bering}
\end{figure}

As shown in Fig. \ref{fig:Bering}, the roll-over is even more apparent here
than in the previous observations due to the larger number of data points and smaller experimental errors.
Overall, the same comments as in the previous section can be made on the comparison
between the measurements and predictions.
The CGG model can somewhat reproduce the roll-over of $q$ near $f = 0.13$ Hz,
despite the fact that the corresponding fit appears smoother in this region.
It underestimates the peak amplitude and slightly overshoots the peak frequency.
Interestingly, these relative features of the roll-over from the field observations and numerical estimates
are reminiscent of the comparison given in \cite{lhv91} (see their Fig. 13)
between their viscoelastic theory and the same Bering Sea data.
Note that attenuation rate is plotted as a function of wave period rather than frequency in their Fig. 13,
where the roll-over takes place at short periods.
At the opposite end of the spectrum, the CGG fit is also found to provide a good approximation for the low-frequency tail.

We see again in Fig. \ref{fig:Bering} that none of the other models produce a roll-over.
The associated curves are all monotonically increasing with frequency, 
and look similar to those in Fig. \ref{fig:Greenland}, although they seem to display a more convex shape here.
This convexity is especially pronounced for the KM and WS curves (the former is extracted from Fig. 11 in \cite{km08}).
Notice again that the KM curve is rougher than the other curves for the same reason as mentioned earlier.

In comparison to these field data, the pore size deduced from the CGG fit turns out to be $a = 22.0$ m, 
which is not so different from the previous prediction ($a = 14.6$ m) for the Greenland Sea experiment.
Given the denser floe field here, we might have expected a smaller pore size,
nevertheless this value $a = 22.0$ m is definitely larger than those found for the frazil/pancake ice covers
generated in the laboratory experiments of Wang and Shen \cite{ws10b} and Zhao and Shen \cite{zs15}.
It is striking how close this estimated pore size is to the floe diameter $\ell = 14.5$ m
that was assumed by Perrie and Hu \cite{ph96} in their simulations of the Bering Sea observations.

Although it is difficult to discriminate any specific physical mechanism from the CGG formulation,
which would be responsible for the observed roll-over, we recall preliminary results from \cite{cgg19}
suggesting that the relative motion between different constituents of the ice cover
induces friction that may interfere with other (bulk) dissipative effects to help produce
this non-monotonic behavior of the attenuation rate.
As can be seen from the definition \eqref{friction} of its controlling parameter $b$, 
this phenomenon in the CGG view is directly linked to the porous (hence heterogeneous) nature of the ice cover.
Details of the frictional process are unclear in this effective medium approach
and so we prefer not to attempt to interpret it further at this point.
We have confirmed the previous presumption by fitting the CGG model to the Bering and Greenland Seas data
in the absence of frictional effects (i.e. with $b$ manually set to zero).
No roll-over has emerged from these computations (this is not shown here for convenience); 
the CGG curve would be monotonically increasing with frequency and would look similar to the WS curve.

\subsection{Field observations of Kohout et al. (2014)}

We turn our attention to a more recent data set that was collected in the Antarctic MIZ 
as part of the Australian Antarctic Division's second Sea Ice Physics and Ecosystem Experiment in 2012 \cite{kw13}.
Wave measurements were made simultaneously using contemporary sensors at up to five locations
on a transect spanning up to $250$ km.
Kohout et al. \cite{kwdm14} provided a preliminary report on these measurements to support the claim 
that wave activity and ice extent are correlated.
A spectral analysis of the data was performed by Meylan et al. \cite{mbk14} who examined in particular the dependence
of attenuation rates on wave periods.

Following Mosig et al. \cite{mms15}, we assume $h = 1$ m and $H = 4300$ m for our computations in this setting.
Four estimates of mean ice concentration $C = 0.210$, $0.481$, $0.498$, $0.576$
in areas of the Antarctic MIZ where the wave sensors drifted, are given in \cite{mbk14}.
These estimates were calculated using Nimbus-7 scanning multichannel microwave radiometer 
and Defense Meteorological Satellite Program (DMSP) Special Sensor Microwave/Imager Sounder (SSMIS) 
Passive Microwave Data.
As a representative value, we specify the average $\beta = 0.56$ of the corresponding ice porosities in the CGG model.
A camera installed on the upper deck of the ship monitored the floe size distribution during this expedition.
Photographs of the broken floe field taken by this camera can be seen in \cite{mbk14}.

\begin{figure}[t!]
\centering
\includegraphics[width=\linewidth]{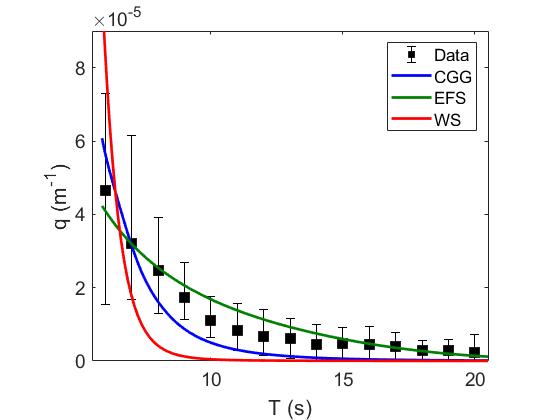}
\caption{Comparison of attenuation rate vs. period between model predictions and data for a broken floe field from the Antarctic MIZ \cite{kwdm14,mbk14}.
Predictions from the CGG, EFS and WS models are shown.
Results for $h = 1$ m and $H = 4300$ m are presented.}
\label{fig:Antarctic}
\end{figure}

Attenuation rates extracted from Fig. 8 in \cite{mms15} (see also Fig. 4 in \cite{mbk14}) are now shown in Fig. \ref{fig:Antarctic} 
and compared to theoretical predictions.
Again, we take into account the difference between data on wave energy decay from \cite{mbk14}
and estimates on wave amplitude decay from the various models by halving the former decay rates.
For this particular data set, we plot $q$ as a function of $T$ (wave period) rather than $f$ (wave frequency),
as originally presented in \cite{mbk14,mms15}, to retain a uniform resolution over the range of periods considered.
Unlike the field observations discussed in the previous section, no roll-over is discernible from the data points
in Fig. \ref{fig:Antarctic}. Accordingly, none of the models involved in this comparison (including the CGG model)
predict such a phenomenon; their fitting curves are all monotonically decreasing with increasing $T$.

Here the EFS model provides the closest fit as indicated in Fig. \ref{fig:Antarctic}.
The agreement is especially good at long periods while, as $T \rightarrow 0$, 
this model tends to underestimate the attenuation rate.
Note that our version of the EFS curve bears a resemblance to the original one shown in \cite{mms15} (see their Fig. 8),
which may be viewed as further evidence for the effectiveness of our fitting procedure.
By contrast, the CGG and WS curves are steeper, falling down more quickly as $T$ increases.
These two models produce negligible values of $q$ at long periods,
which are distinctly lower than the field data over most of the time interval being probed.
On the other hand, they tend to overestimate the decay rates at short periods.
Despite these discrepancies, the CGG fit is seen to lie within or near experimental error,
while the WS fit tends to lie further below.
We remark in passing that the decay rates observed in this case and in the Arctic MIZ \cite{wsgcm88}
are significantly lower than those measured in the laboratory experiments as discussed previously.
This supports a previous statement from Section 2 that the decay rates in frazil/pancake ice can be
several orders of magnitude greater than in a broken floe field 
(e.g. compare values of $q$ between Figs. \ref{fig:WangShen} and \ref{fig:Antarctic}).

The pore size returned by the CGG fit to these field measurements is $a = 72.0$ m,
which is larger than the predictions for the two previous data sets from the Arctic MIZ.
While the photographs in \cite{mbk14} might suggest a lower value of $a$ for this broken floe field,
we point out that these were taken immediately after deployment of the sensors.
Over the duration of their operation, these sensors tended to drift into open ocean, as mentioned in \cite{mbk14}.
Furthermore, considering that the MIZ explored was overall more on the sparse side ($\beta = 0.56$),
with dominant floe sizes $\ell$ ranging from a few meters to greater than $100$ m
in the various areas visited by the sensors, we deem the pore size $a = 72.0$ m 
estimated from the CGG model to be reasonable here as well.

\section{Discussion on shear modulus and kinematic viscosity}

We further check the performance of the CGG model by comparing its estimates of shear modulus $\mu$ 
and kinematic viscosity $\eta$ with predictions by the EFS and WS models.
These two parameters are important measures of effective viscoelastic properties of the ice cover
and are common to all three continuum formulations.
Such an assessment would be suitable as part of the calibration of these models in view of potential applications
to large-scale wave forecasting in the polar regions.
Similar parametric calibration of the EFS and WS models has been conducted in \cite{cheng17,mms15,zs15},
although these studies used different methods to fit the theoretical predictions to experimental data.

Table 1 lists values of $\mu$ and $\eta$ as determined from the CGG, EFS and WS fits
to the laboratory experiments and field observations that we discussed in the previous sections.
To highlight the effective character of these models as applied to wave propagation in various ice-cover types,
these estimates are presented in such a way that they are normalized relative to typical values
$\mu \simeq \mu_0 = 10^9$ Pa for pack ice \cite{mms15,wf92} 
and $\eta \simeq \eta_0 = 10^{-2}$ m$^2$ s$^{-1}$ for grease or pancake ice \cite{ddmbw15,nm99}.
As alluded to in Section 2, Table 1 confirms that both parameters can take a wide range of values
depending on the particular model and ice conditions.
Among all the cases considered, the highest values of $\mu$ and $\eta$ are both achieved by the EFS fit
to the Greenland Sea observations \cite{wsgcm88}.
The lowest value of $\mu$ is given by the WS fit to the laboratory Test 1 of Newyear and Martin \cite{nm97}.
The lowest value of $\eta$ is returned by the CGG fit to the Tank 2 experiment of Wang and Shen \cite{ws10b}.
These extreme values are highlighted in blue (highest) and red (lowest) in Table 1.
As expected, for all three models, the shear modulus $\mu$ (which is a measure of the ice-cover's elasticity)
is found to be smallest for grease ice (Test 1 in \cite{nm97}) and largest for a broken floe field (Arctic MIZ \cite{wsgcm88}).
Their estimates of $\mu$ for both Antarctic and Arctic MIZ remain overall within two orders of magnitude
from the typical value $\mu_0$ for pack ice.
By contrast, the kinematic viscosity $\eta$ (which may represent a combination of various attenuating effects 
in this continuum framework) exhibits a more complicated behavior depending on the particular model and ice conditions.
We point out however that all three models predict $\eta$ to be on the order of $\eta_0$
for the grease-ice experiments \cite{nm97}, which is consistent with values $\eta = 2$--$3 \, \eta_0$
inferred by Newyear and Martin \cite{nm99} who fitted data from \cite{nm97} to Keller's two-layer viscous model \cite{k98}.
For a broken floe field, the viscosity estimates from both EFS and WS models are found to be larger 
by several orders of magnitude than their counterparts for grease ice.
On the other hand, the corresponding predictions from the CGG model remain comparable between these two types of ice cover.

On a related note, we see that $\mu$ and $\eta$ as determined by the CGG fit only vary over $6$ and $3$ orders of magnitude
respectively, among all the cases considered.
By contrast, $\mu$ and $\eta$ as predicted by the WS fit vary over $10$ and $6$ orders of magnitude respectively,
while both parameters in the EFS model vary over $10$ orders of magnitude.
This suggests that both $\mu$ and $\eta$ in the CGG model may only require moderate tuning 
in view of potential applications to operational wave forecasting.
Our estimates of $\mu$ ($4.2 \times 10^{11}$ Pa) and $\eta$ ($4.2 \times 10^6$ m$^2$ s$^{-1}$) from the EFS fit
to the Antarctic MIZ data are consistent with those reported in \cite{mms15} for the same model 
($\mu = 4.9 \times 10^{12}$ Pa, $\eta = 5.0 \times 10^7$ m$^2$ s$^{-1}$).
Both of them are several orders of magnitude larger than the reference values $\mu_0$ and $\eta_0$ (especially for $\eta$).
Regardless of how close the fit is, the EFS model tends to require very large values of these parameters
in order to reproduce wave attenuation in broken floe fields of the Antarctic and Arctic MIZ.
This may be interpreted as a way to make up for the thin-plate approximation so that elastic properties
of the ice cover would be sufficiently well captured in these situations.

Our estimates of $\mu$ ($\{ 4.2, 2.5 \times 10^5, 8.3 \times 10^5 \}$ Pa) and 
$\eta$ ($\{ 1.5 \times 10^{-2}, 45.0, 131.6 \}$ m$^2$ s$^{-1}$) from the WS fit to the laboratory measurements
of Zhao and Shen \cite{zs15} are in good agreement with their own findings ($\mu = \{ 21, 5 \times 10^5, 1 \times 10^6 \}$ Pa,
$\eta = \{ 1.4 \times 10^{-2}, 61, 140 \}$ m$^2$ s$^{-1}$) for Test 1, 2, 3 respectively (see their Table 2).
These authors also fitted the WS model to a data set from \cite{ws10b} (it was not clearly stated which experiment was considered)
and obtained $\mu = 48$ Pa, $\eta = 4 \times 10^{-2}$ m$^2$ s$^{-1}$ which again are fairly close 
in terms of order of magnitude to our own results ($\mu = \{ 1.1 \times 10^3, 4.0 \times 10^2 \}$ Pa, 
$\eta = \{ 9.6 \times 10^{-2}, 5.1 \times 10^{-2} \}$ m$^2$ s$^{-1}$) for Tank 2 and 3 respectively.
When examining the CGG and WS models against the field observations, 
we see that their predictions of $\mu$ are comparable to each other on the order of $10^7$ Pa,
which contrasts with the much higher values from the EFS fit, as noted above.
This similarity however does not extend to $\eta$ since the WS model yields values that are higher than
the CGG predictions by several orders of magnitude.
Again, the range of estimated $\eta$ from the CGG fit is strikingly narrow among all the cases considered,
in comparison to the other two models.

It is also worth mentioning that the estimates $\mu = 3.3 \times 10^7$ Pa and $\eta = 1.2 \times 10^{-2}$ m$^2$ s$^{-1}$  
from the CGG fit to the Bering Sea measurements are consistent with those 
($\mu = 2.3 \times 10^9$ Pa, $\eta = 1.5 \times 10^{-2}$ m$^2$ s$^{-1}$) reported in \cite{lhv91} for the same data set.
As stated earlier, these authors used a thin-plate viscoelastic model and were able to emulate the roll-over phenomenon
to some extent (see their Fig. 13).
In that study, $\mu$ was assigned a typical value $\sim \mu_0$ for pack ice while $\eta$ was deduced from the data fitting.
Interestingly, the fitting curve shown in Fig. 13 of \cite{lhv91} bears a resemblance to the CGG curve 
in our Fig. \ref{fig:Bering} (modulo the switch between wave frequency and period for the horizontal axis).
Lastly, we remark that the CGG predictions of $\eta \sim 10^{-2}$--$10^0 \, \eta_0$ for the data sets
from the Antarctic and Arctic MIZ are encouraging in view of earlier measurements that reported values of eddy viscosity 
under large ice floes, ranging from $2.4 \times 10^{-3}$ m$^2$ s$^{-1}$ in the central Arctic Ocean \cite{h66}
to $2.1 \times 10^{-2}$ m$^2$ s$^{-1}$ in the Weddell Sea (Antarctic MIZ) \cite{mm94}.

\section{Sensitivity tests}

Given the rather large number of rheological parameters associated with the ice cover in the CGG formulation,
it is of interest to check their individual relevance to this problem and test the sensitivity of attenuation rate predictions
with respect to these parameters.
For this purpose, we take the Bering Sea observations as a representative discriminating case
because it exhibits unusual features such as the roll-over phenomenon and contains a fair number of data points.
We focus our attention on the following parameters: $h$ (thickness), $\beta$ (porosity), $a$ (pore size),
$\eta$ (kinematic viscosity), $\mu$ (shear modulus) and $\nu$ (Poisson's ratio).
As alluded to in previous sections, some of them which are related to geometrical features of the ice cover
(e.g. thickness, porosity, pore size) may be estimated by in-situ measurement or remote sensing,
while others which are related to material properties (e.g. kinematic viscosity, shear modulus)
would be more difficult to determine or guess.
With this in mind, a sensitivity analysis may help assign predefined values to some of these parameters
(as opposed to other parameters that may require more tuning), in order to reduce the parameter space for the CGG model.

\begin{figure}[t!] 
\centering
\begin{subfigure}{0.52\textwidth}
\includegraphics[width=\linewidth]{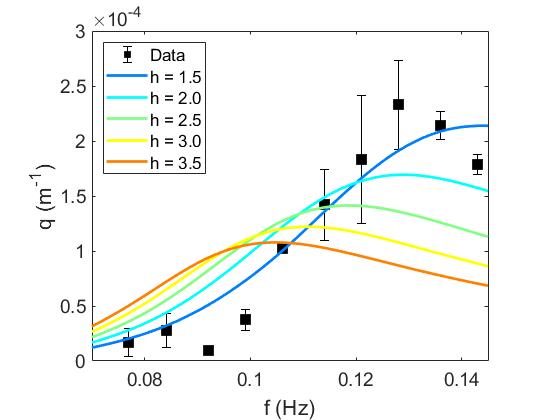}
\caption{varying thickness $h$} \label{fig:h}
\end{subfigure}\hspace*{\fill}
\begin{subfigure}{0.52\textwidth}
\includegraphics[width=\linewidth]{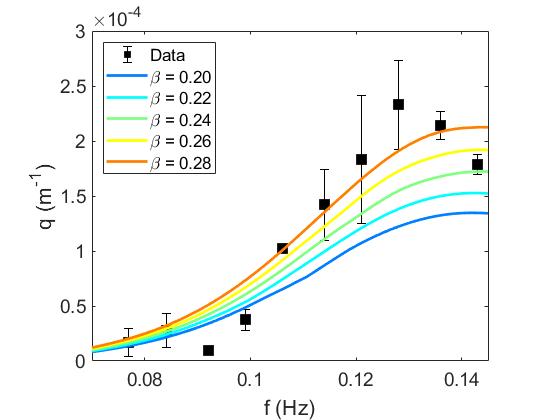}
\caption{varying porosity $\beta$} \label{fig:beta}
\end{subfigure}
\begin{subfigure}{0.52\textwidth}
\includegraphics[width=\linewidth]{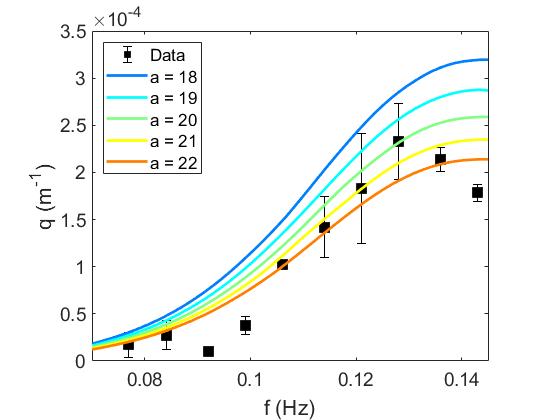}
\caption{varying pore size $a$} \label{fig:a}
\end{subfigure}\hspace*{\fill}
\begin{subfigure}{0.52\textwidth}
\includegraphics[width=\linewidth]{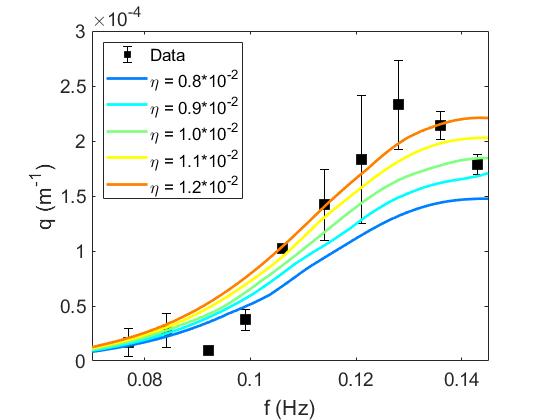}
\caption{varying kinematic viscosity $\eta$} \label{fig:eta}
\end{subfigure}
\begin{subfigure}{0.52\textwidth}
\includegraphics[width=\linewidth]{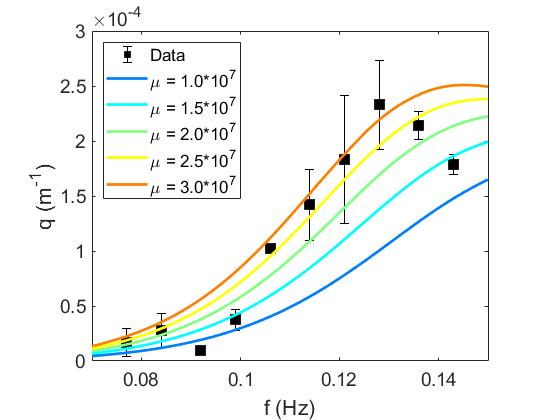}
\caption{varying shear modulus $\mu$} \label{fig:mu}
\end{subfigure}\hspace*{\fill}
\begin{subfigure}{0.52\textwidth}
\includegraphics[width=\linewidth]{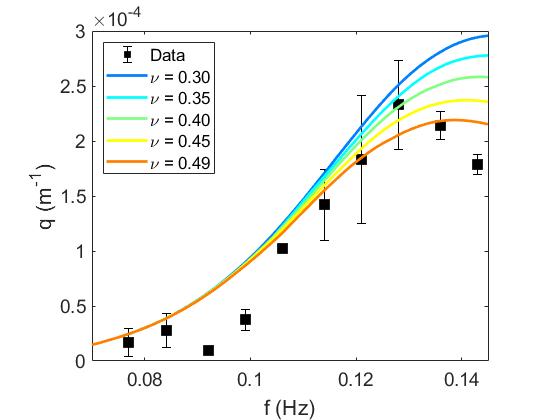}
\caption{varying Poisson's ratio $\nu$} \label{fig:nu}
\end{subfigure}
\caption{Sensitivity of attenuation rate vs. frequency to varying parameters as predicted by the CGG model.
Data for a broken floe field from the Bering Sea 7 February 1983 experiment \cite{wsgcm88} are considered.
Results for varying (a) thickness $h$ (m), (b) porosity $\beta$, (c) pore size $a$ (m), (d) kinematic viscosity $\eta$ (m$^2$ s$^{-1}$),
(e) shear modulus $\mu$ (Pa), (f) Poisson's ratio $\nu$ are presented.} 
\label{fig:parameters}
\end{figure}

For each of these parameters, Fig. \ref{fig:parameters} displays a set of curves for the attenuation rate 
as predicted by the CGG model. The reference regime of parameters is given by the corresponding best fit to the Bering Sea data,
as discussed in the previous section (see Fig. \ref{fig:Bering}).
This set of curves is obtained by varying the parameter under consideration
while freezing the other parameters at their original best-fitting values.
The objective of such an analysis is to examine how perturbations in individual parameters would affect the original best fit.
The range of perturbations for each parameter is chosen to be an interval around its best-fitting value.

Ice thickness is a distinctive feature of the ice cover in the CGG formulation, 
as opposed to the viewpoint in the thin-plate approximation.
Fig. \ref{fig:parameters}(a) reveals that the roll-over tends to shift upward and to higher frequencies as $h$ is decreased.
This tendency is quite pronounced and suggests strong sensitivity of $q$ with respect to $h$.
A decrease in $h$ by a factor of $3$ shifts the peak outside the frequency range of the experimental data,
and moves it out of sight in this figure.
The fact that shorter waves (i.e. at higher frequencies) experience more attenuation in thinner nice (i.e. for smaller $h$),
which is rather counter-intuitive, is reminiscent of a common feature in models for water waves over seabed
composed of a viscous mud layer, where dissipation has a non-monotonic dependence on mud-layer thickness,
with thicker layers being less dissipative \cite{cgg19,crl17,dl78}.
Ice porosity and pore size are rheological parameters that are characteristic of the present model.
As illustrated in Figs. \ref{fig:parameters}(b) and (c), increasing $\beta$ or decreasing $a$ has the basic effect of raising the attenuation rate
and accentuating the roll-over. In either case, the peak remains around $f = 0.13$ Hz as $\beta$ or $a$ is varied.
Note that the dichotomy in variation between $\beta$ and $a$ is attributed to their contrasting roles in the friction process. 
Because the parameter $b$ depends linearly on $\beta$ while it is inversely proportional to $a^2$ according to \eqref{friction}, 
friction is thus enhanced (and so is the roll-over) as $\beta$ is increased or $a$ is decreased.
A similar behavior occurs as $\eta$ is increased (see Fig. \ref{fig:parameters}d), 
which is anticipated considering the linear dependence of $b$ on $\eta$.
A slightly more complicated picture is observed for the variation with respect to $\mu$.
Inspecting Fig. \ref{fig:parameters}(e), the roll-over tends to shift upward and to lower frequencies as $\mu$ is increased.
The sensitivity of $q$ to $\nu$ is relatively weak and is confined to the high-frequency region, as suggested by Fig. \ref{fig:parameters}(f).
This explains why, for convenience and given that $0 < \nu < 1/2$, we set $\nu = 0.4$ in the previous computations
(which is close to the typical value $\nu = 1/3$ for pack ice \cite{wf92}).

Our sensitivity tests indicate that all these parameters have some influence on the roll-over,
affecting its amplitude and/or position.
Sensitivity of $q$ with respect to $h$ and $\mu$ seems to be most nontrivial, and is particularly strong for $h$.
Liu et al. \cite{lhv91} also concluded from their model-data comparison that the frequency at which the roll-over occurs
depends on ice conditions, especially ice thickness.
In light of this sensitivity analysis and results from Section 3, to help reduce the parameter space,
it would also be reasonable to fix the pore size with some predefined value of order $O(10)$ m
for potential applications to wave forecasting in the MIZ.
This is even more relevant considering that this parameter only appears in the expression \eqref{friction} 
of the friction coefficient for the CGG model.

\section{Conclusions}

To assess the recently proposed CGG model, we test it against a selection of laboratory experiments
and field observations taken from the literature, concerning wave attenuation in sea ice.
Altogether, these measurements span a wide range of ice conditions and wave frequencies.
We fit the theoretical predictions to data on attenuation rate via error minimization,
which in turn yields estimates for effective rheological parameters
in addition to estimates for the attenuation rate.
Whenever the information is available, the porosity parameter is assigned a value that is the complement of the mean ice concentration.
To further check this model's performance, we also compare it to other existing viscoelastic theories
under the same various conditions.
Numerical solutions of the dispersion relations can be found using relatively simple selection criteria.
As a byproduct, we independently recover (via a different fitting procedure) a number of results 
that are similar to those reported in previous studies.
Special attention is paid to the EFS and WS formulations which share some common features with the CGG system.
For such parameters as $\mu$ (shear modulus) and $\eta$ (kinematic viscosity) which control the viscoelastic properties, 
we find that the range of estimated values (over all the situations considered) may differ significantly from one model to another.
Among these three representations, the CGG (resp. EFS) model turns out to be the one for which 
the predicted range of both $\mu$ and $\eta$ is the narrowest (resp. widest) in orders of magnitude.
Even for individual situations, this difference in parameter recovery may be considerable.
As expected, for grease ice, all three models predict $\eta$ on the order of $\eta_0 = 10^{-2}$ m$^2$ s$^{-1}$
and $\mu$ to be essentially negligible compared to the typical value $\mu_0 = 10^9$ Pa for pack ice.
On the other hand, for broken floe fields, there is more variability in the determination of these parameters, especially for $\eta$.
We obtain in this case values of $\mu$ that may be lower (for CGG and WS) or higher (for EFS) 
than $\mu_0$ by a few orders of magnitude.
Estimates of $\eta$ from the EFS and WS fits tend to be larger than $\eta_0$ 
by several orders of magnitude, while those from the CGG fit remain around this reference value.
Overall, the CGG model provides good fits to the data on attenuation rate for the various cases under consideration.
Against the Antarctic MIZ data, the EFS counterpart appears to be a clear favorite,
but the corresponding fit is achieved for values of $\mu$ and $\eta$ that are both excessively high,
a fact which has also been pointed out in \cite{mms15}.
By contrast, the CGG fit returns markedly lower values for these parameters, with $\mu$ being lower than $\mu_0$
by two orders of magnitude and $\eta$ being essentially equal to $\eta_0$.
In comparison to the Arctic MIZ data (from both the Bering and Greenland Seas), 
the CGG model is able to reasonably reproduce the roll-over of attenuation rate,
unlike both EFS and WS counterparts which only predict a monotonic growth with frequency.
According to the poroelastic formulation, this intriguing phenomenon is attributed to friction caused by
the relative motion between fluid and solid components of the ice cover,
which highlights the role of porosity in the present description of wave-ice interactions,
as such friction is directly connected to the porous nature of the ice cover.
This dissipative mechanism could possibly be a contributing factor in the roll-over effect as reported in field observations.

In the future, various extensions of this work may be envisioned.
Further calibration of the CGG model would be suitable using more data (and larger data sets),
which may include e.g. recent data from the Arctic MIZ as in \cite{cheng17}.
Using such data requires substantial processing and analysis.
It would also be of interest to perform more tests against other data sets exhibiting a roll-over
in order to further assess the possible role of friction in this phenomenon.
While the CGG system features a larger number of physical parameters than other existing viscoelastic formulations.
this study suggests that some of these parameters may be assigned predefined values,
or may be estimated by in-situ measurement or remote sensing.
Moreover, it is conceivable that the parametric dependence in the CGG model may be mathematically simplified
by examining asymptotic regimes (in the limit of vanishing parameters) according to the specific type of ice cover under consideration.
Finally, it would be appropriate to extend these results to the three-dimensional setting
(for wave propagation in two horizontal directions) as well as to the nonlinear case.
Discrepancies that we have observed in comparison to experimental data may partly be attributed to such effects.
Nonlinear theory of wave-ice interactions has drawn increasing attention in recent years \cite{dkp19,gp12,gp14,gp17}.

\clearpage

\begin{table}[!t]
\hspace{-4cm}
\begin{tabular}{|c|c|c|c|c|c|c|c|}
\hline
\multirow{3}{*}{Experiment} & \multirow{3}{*}{Data set} & \multicolumn{2}{c|}{CGG model} & \multicolumn{2}{c|}{EFS model} & \multicolumn{2}{c|}{WS model} \\
\cline{3-8}
& & $\mu$ ($\times 10^9$) & $\eta$ ($\times 10^{-2}$) & $\mu$ ($\times 10^9$) & $\eta$ ($\times 10^{-2}$) & $\mu$ ($\times 10^9$) & $\eta$ ($\times 10^{-2}$) \\
& & (Pa) & (m$^2$ s$^{-1}$) & (Pa) & (m$^2$ s$^{-1}$) & (Pa) & (m$^2$ s$^{-1}$) \\
\hline
\multirow{2}{*}{Newyear \& Martin} & Test 1 & $5.00 \times 10^{-7}$ & $7.52$ & $1.17 \times 10^{-7}$ & $2.50$ & \color{red}{$6.40 \times 10^{-11}$} & $2.80$ \\
& Test 2 & $6.00 \times 10^{-7}$ & $9.00$ & $1.26 \times 10^{-7}$ & $2.64$ & $1.20 \times 10^{-10}$ & $3.76$ \\
\hline
\multirow{2}{*}{Wang \& Shen} & Tank 2 & $1.82 \times 10^{-4}$ & \color{red}{$1.22 \times 10^{-2}$} & $4.80 \times 10^{-4}$ & $2.25 \times 10^4$ & $1.15 \times 10^{-6}$ & $9.61$ \\
& Tank 3 & $1.73 \times 10^{-4}$ & $4.00 \times 10^{-2}$ & $2.80 \times 10^{-5}$ & $1.20 \times 10^3$ & $4.00 \times 10^{-7}$ & $5.10$ \\
\hline
\multirow{3}{*}{Zhao \& Shen} & Test 1 & $7.34 \times 10^{-6}$ & $8.32 \times 10^{-2}$ & $7.20 \times 10^{-6}$ & $8.00 \times 10^2$ & $4.20 \times 10^{-9}$ & $1.46$ \\
& Test 2 & $4.21 \times 10^{-5}$ & $4.68 \times 10^{-2}$ & $9.40 \times 10^{-4}$ & $1.62 \times 10^4$ & $2.47 \times 10^{-4}$ & $4.50 \times 10^3$ \\
& Test 3 & $1.44 \times 10^{-4}$ & $9.22 \times 10^{-2}$ & $7.20 \times 10^{-2}$ & $1.44 \times 10^6$ & $8.32 \times 10^{-4}$ & $1.32 \times 10^4$ \\
\hline
\multirow{2}{*}{Wadhams et al.} & Greenland Sea & $1.38 \times 10^{-2}$ & $7.00 \times 10^{-2}$ & \color{blue}{$6.50 \times 10^2$} & \color{blue}{$4.62 \times 10^9$} & $6.75 \times 10^{-2}$ & $1.14 \times 10^5$ \\
& Bering Sea & $3.30 \times 10^{-2}$ & $1.16$ & $1.54$ & $5.28 \times 10^6$ & $4.00 \times 10^{-2}$ & $2.00 \times 10^5$ \\
\hline
Kohout et al. & Antarctic MIZ & $1.58 \times 10^{-2}$ & $1.00$ & $4.20 \times 10^2$ & $4.20 \times 10^8$ & $6.00 \times 10^{-3}$ & $1.20 \times 10^4$ \\
\hline
\end{tabular}
\caption{Estimates of shear modulus and kinematic viscosity from the CGG, EFS and WS fits 
to data on attenuation rate for the various cases under consideration.
Values of shear modulus are normalized relative to $\mu_0 = 10^9$ Pa, 
while values of kinematic viscosity are normalized relative to $\eta_0 = 10^{-2}$ m$^2$ s$^{-1}$.
Lowest estimates are highlighted in red while highest estimates are highlighted in blue.}
\end{table}

\clearpage

\appendix

\section{Coefficients in the dispersion relation}

Coefficients in the dispersion relation \eqref{poro-disp} have the following expressions:
\begin{equation*}
\begin{aligned}
T_1 =& \, 2 E_{1} E_{12} [E_{2} E_{23} (1 - \cosh(D_{1} h) \cosh(D_{3} h)) \sinh(D_{2} h)  \\
& -E_{1} E_{13} (1 - \cosh(D_{1} h) \cosh(D_{2} h)) \sinh(D_{3} h)]   \\
& +\sinh(D_{1} h) [2 E_{1} E_{13} E_{2} E_{23} (-1 +  \cosh(D_{2} h) \cosh(D_{3} h))  \\
& -(E_{1}^2 (E_{12}^2 + E_{13}^2) + E_{2}^2 E_{23}^2) \sinh(D_{2} h) \sinh(D_{3} h)] ~,\\
T_2 =& \,E_{1} [(E_{17} E_{4} E_{7} + E_{18} E_{6} E_{9}) (1 - \cosh(D_{1} h) \cosh(D_{2} h)) \\
& - (E_{14} E_{2} E_{21} E_{7} - E_{20} E_{5} E_{9}) (1 - \cosh(D_{1} h) \cosh(D_{3} h))  \\
& - (E_{19} E_{4} E_{5} - E_{14} E_{2} E_{22} E_{6}) (1 - \cosh(D_{2} h) \cosh(D_{3} h))]  \\
& - E_{1} [(E_{17} E_{3} E_{7} + E_{18} E_{6} E_{8}) (1 - \cosh(D_{1} h) \cosh(D_{2} h)) \cosh(D_{3} h)  \\
& + (E_{16} E_{2} E_{21} E_{7} - E_{20} E_{5} E_{8}) \cosh(D_{2} h) (-1 + \cosh(D_{1} h) \cosh(D_{3} h))  \\
& + (E_{19} E_{3} E_{5} - E_{16} E_{2} E_{22} E_{6}) \cosh(D_{1} h) (-1 + \cosh(D_{2} h) \cosh(D_{3} h))]  \\
& + E_{1} (E_{12} E_{4} E_{6} + E_{13} E_{7} E_{9}) \sinh(D_{1} h) \sinh(D_{2} h)  \\
& - E_{1} (E_{12} E_{3} E_{6} + E_{13} E_{7} E_{8}) \cosh(D_{3} h) \sinh(D_{1} h) \sinh(D_{2} h)  \\
& + (E_{1}^2 E_{12} E_{14} E_{5} - E_{2} E_{23} E_{7} E_{9}) \sinh(D_{1} h) \sinh(D_{3} h)  \\
& - (E_{1}^2 E_{12} E_{16} E_{5} - E_{2} E_{23} E_{7} E_{8}) \cosh(D_{2} h) \sinh(D_{1} h) \sinh(D_{3} h)  \\
& - (E_{1}^2 E_{13} E_{14} E_{5} - E_{2} E_{23} E_{4} E_{6}) \sinh(D_{2} h) \sinh(D_{3} h)  \\
& + (E_{1}^2 E_{13} E_{16} E_{5} - E_{2} E_{23} E_{3} E_{6}) \cosh(D_{1} h) \sinh(D_{2} h) \sinh(D_{3} h) ~,
\end{aligned}
\end{equation*}
\begin{equation*}
\begin{aligned}
T_3 =&\, E_{1} [(E_{11} E_{18} E_{6} + E_{10} E_{17} E_{7}) (1 - \cosh(D_{1} h) \cosh(D_{2} h)) \cosh(D_{3} h)  \\
& - (E_{11} E_{20} E_{5} - E_{15} E_{2} E_{21} E_{7}) \cosh(D_{2} h) (-1 + \cosh(D_{1} h) \cosh(D_{3} h))  \\
& + (E_{10} E_{19} E_{5} - E_{15} E_{2} E_{22} E_{6}) \cosh(D_{1} h) (-1 + \cosh(D_{2} h) \cosh(D_{3} h))]  \\
& + E_{1} (E_{10} E_{12} E_{6} + E_{11} E_{13} E_{7}) \cosh(D_{3} h) \sinh(D_{1} h) \sinh(D_{2} h)  \\
& + (E_{1}^2 E_{12} E_{15} E_{5} - E_{11} E_{2} E_{23} E_{7}) \cosh(D_{2} h) \sinh(D_{1} h) \sinh(D_{3} h)  \\
& - (E_{1}^2 E_{13} E_{15} E_{5} - E_{10} E_{2} E_{23} E_{6}) \cosh(D_{1} h) \sinh(D_{2} h) \sinh(D_{3} h) ~,
\end{aligned}
\end{equation*}
and the $E_{j} \, (j = 1, \dots, 23)$ are defined as follows:
\begin{equation*}
\begin{aligned}
E_{1} =&\,  2 D_{3} \kappa^2 \mu_c ~, \\
E_{2} =&\, -(D_{2}^2 - \kappa^2) (F_{6} Q + F_{8} R) (-\kappa^2 (F_{7} Q + F_{5} \lambda) +  D_{1}^2 (F_{7} Q + F_{5} (\lambda + 2 \mu_c)))  \\
& + (D_{1}^2 - \kappa^2) (F_{5} Q + F_{7} R) (-\kappa^2 (F_{8} Q + F_{6} \lambda) + D_{2}^2 (F_{8} Q + F_{6} (\lambda + 2 \mu_c))) ~, \\
E_{3} =&\, -(D_{2}^2 - \kappa^2) (F_{6} Q + F_{8} R) (\rho_{s} - \rho_{f}) (1 - \beta) ~, \\
E_{4} =&\, -(D_{2}^2 - \kappa^2) (F_{6} Q + F_{8} R) \rho_{s} (1 - \beta)  \\
& + \rho_{f} \beta [-\kappa^2 (F_{8} Q + F_{6} \lambda) + D_{2}^2 (F_{8} Q + F_{6} (\lambda + 2 \mu_c))] ~, \\
E_{5} =&\, -2 D_{1} D_{2} F_{6} (F_{5} (-1 + \beta) - F_{7} \beta) + 
   2 D_{1} D_{2} F_{5} (F_{6} (-1 + \beta) - F_{8} \beta) ~, \\
E_{6} =&\, D_{1} (D_{3}^2 + \kappa^2) (F_{5} (-1 + \beta) - F_{7} \beta) + 
   2 D_{1} F_{5} \kappa^2 (1 + (-1 + F9) \beta) ~, \\
E_{7} =&\, D_{2} (D_{3}^2 + \kappa^2) (F_{6} (-1 + \beta) - F_{8} \beta) + 
   2 D_{2} F_{6} \kappa^2 (1 + (-1 + F9) \beta) ~, \\
E_{8} =&\, -(D_{1}^2 - \kappa^2) (F_{5} Q + F_{7} R) (\rho_{s} - \rho_{f}) (1 - \beta) ~, \\
E_{9} =&\, -(D_{1}^2 - \kappa^2) (F_{5} Q + F_{7} R) \rho_{s} (1 - \beta)  \\
& + \rho_{f} \beta [-\kappa^2 (F_{7} Q + F_{5} \lambda) + D_{1}^2 (F_{7} Q + F_{5} (\lambda + 2 \mu_c))] ~, \\
E_{10} =&\, -(D_{2}^2 - \kappa^2) (F_{6} Q + F_{8} R) \rho_{f} (1 - \beta)  \\
& + \rho_{f} \beta [-\kappa^2 (F_{8} Q + F_{6} \lambda) + D_{2}^2 (F_{8} Q + F_{6} (\lambda + 2 \mu_c))] ~, \\
E_{11} =&\, -(D_{1}^2 - \kappa^2) (F_{5} Q + F_{7} R) \rho_{f} (1 - \beta)  \\
& + \rho_{f} \beta [-\kappa^2 (F_{7} Q + F_{5} \lambda) + D_{1}^2 (F_{7} Q + F_{5} (\lambda + 2 \mu_c))] ~, \\
E_{12} =&\, 2 D_{1} F_{5} (D_{2}^2 - \kappa^2) (F_{6} Q + F_{8} R) ~, \\
E_{13} =&\, 2 D_{2} F_{6} (D_{1}^2 - \kappa^2) (F_{5} Q + F_{7} R) ~, \\
E_{14} =&\, \rho_{f} \beta ~, \\
E_{15} =&\, \rho_{f} \beta ~, \\
E_{16} =&\, 0 ~, \\
E_{17} =&\, 2 D_{1} F_{5} (D_{1}^2 - \kappa^2) (F_{5} Q + F_{7} R) ~, \\
E_{18} =&\, 2 D_{2} F_{6} (D_{2}^2 - \kappa^2) (F_{6} Q + F_{8} R) ~, \\
E_{19} =&\, (D_{1}^2 - \kappa^2) (D_{3}^2 + \kappa^2) (F_{5} Q + F_{7} R) ~, \\
E_{20} =&\, (D_{2}^2 - \kappa^2) (D_{3}^2 + \kappa^2) (F_{6} Q + F_{8} R) ~, \\
E_{21} =&\, 2 D_{1} F_{5} ~, \\
E_{22} =&\, 2 D_{2} F_{6} ~, \\
E_{23} =&\, D_{3}^2 + \kappa^2 ~.
\end{aligned}
\end{equation*}
The expressions of $\lambda$, $Q$, $R$, $D_j \, (j = 1, 2, 3)$ and $F_j \, (j = 1, \dots, 8)$ 
in terms of wave parameters and rheological parameters can be found in \cite{cgg19}.
The complex shear modulus $\mu_c$ is given by \eqref{mu}.

\end{document}